# Control Flow vs. Data Flow in Distributed Systems Integration: Revival of Flow-based Programming for the Industrial Internet of Things

Wilhelm Hasselbring, Maik Wojcieszak, Schahram Dustdar

Abstract

When we consider the application layer [1] of networked infrastructures, data and control flow are important concerns in distributed systems integration. Modularity is a fundamental principle in software design [2], in particular for distributed system architectures. Modularity emphasizes high cohesion of individual modules and low coupling between modules. Microservices are a recent modularization approach with the specific requirements of independent deployability and, in particular, decentralized data management [3]. Cohesiveness of microservices goes hand-in-hand with loose coupling, making the development, deployment, and evolution of microservice architectures flexible and scalable [4]. However, in our experience with microservice architectures, interactions and flows among microservices are usually more complex than in traditional, monolithic enterprise systems, since services tend to be smaller and only have one responsibility, causing collaboration needs. We suggest that for loose coupling among microservices, explicit control-flow modeling and execution with central workflow engines should be avoided on the application integration level. On the level of integrating microservices, data-flow modeling should be dominant. Control-flow should be secondary and preferably delegated to the microservices. We discuss coupling in distributed systems integration and reflect the history of business process modeling with respect to data and control flow. To illustrate our recommendations, we present some results for flow-based programming in our Industrial DevOps project Titan, where we employ flow-based programming for the Industrial Internet of Things.

Data flow is concerned about where data are routed through a program/system and what transformations are applied during that journey. Control flow is concerned about the possible order of operations. These two concepts are somehow linked to each other: E.g., the order of operations executed in a computer program can influence where the data go. Similar, specific data values may steer the control flow. The side box below discusses data vs. control flow in program analysis as an area related to this article.

---

**Data vs. Control Flow Program Analysis**

In programming, when calling a function, starting the function's execution is control flow while passing the function's parameters is data flow. In this context, control and data flow are tightly linked, thus it is not straight forward to separate them:

- Control-flow analysis deconstructs the order of operations in a computer program. This could be, for example, determining execution paths, but also precedence constraints between different operations.

    The dominant question is how the locus of control moves through the program. Data may accompany the control flow, but is not dominant.

- Data-flow analysis gathers information about the possible set of values calculated at various locations in a computer program.

    The dominant question is how data moves through computations. As the data moves, control is activated.

Control flow refers to the path the execution takes in a program, and sequential programming that focuses on explicit control flow using control structures like loops or conditional statements is called **imperative programming**. In an imperative model, data may follow the control flow, but the main question is about the order of execution.

Dataflow abstracts over explicit control flow by placing the emphasis on the routing and transformation of data and is part of the **declarative programming** paradigm. In a dataflow model, control follows data and computations are executed implicitly based on data availability. Concurrency control refers to the use of explicit mechanisms like locks to synchronize interdependent concurrent computations. It is a matter of emphasis – control flow schedules data movement, or data movement implies transfer of control.

---

# Distributed Systems Integration

File transfer, shared databases and Web resources, remote procedure calls, asynchronous messaging, and data streaming support different forms of dataflow and control flow across distributed integrated systems, with various degrees of coupling introduced in the integrated architecture [5].

Service Oriented Architecture (SOA) is an approach to developing enterprise systems by loosely coupling interoperable services from separate systems across different business domains. SOA emerged in the early 2000s, offering a way to develop new business services by reusing components from existing programs within the enterprise rather than writing functionally redundant code from scratch.

A crucial aspect of SOA is service orchestration. Developers utilize service orchestration to support the automation of business processes. Service orchestration is the coordination of multiple services exposed as a single aggregate service. In other words, service orchestration is the combination of service interactions to create higher-level business services. This is usually accomplished through the use of a central workflow engine and/or an enterprise service bus (ESB). However, such a central orchestration service causes problems in microservice architectures:

- A team should have full-stack responsibility for its microsevices (Conway's Law). With a central orchestration service comes a central workflow team, which has to coordinate with the microservices teams whose services are involved in the workflow. With microservice architectures you create at least one microservice per bounded context, according to domain-driven design [6]. One important goal of microservices is to improve scalability and speed of the software development itself [4]. Hence it is common sense that one microservice needs to be owned by exactly one development team (which may own multiple microservices). Centrally managed ESBs do not fit into a microservices architecture. You may face a situation where you have to update your microservices in-sync with the central workflow model in case you make changes. This introduces a coupling between central and local control, that you do not want to have. However, inside a microservice, or if a team owns multiple microservices, a workflow engine may be appropriate within this context.

- For independent scalability, microservices should manage their data themselves; thus, manage their bounded context. Long running workflows need to keep persistent state somehow. This may imply coupling between the central orchestration service and the individual microservices. With microservices, workflows should only live inside service boundaries, if loose coupling is pursued.

Thus, we observe a conflict between central control (via orchestration) and independent evolution of microserives. This is particularly the case in our application domain of industrial automation, where for instance sensors and actuators are managed by third parties, with highly varying update cycles.

Service choreography is also related to service orchestration, as both are employed to create composite services and applications in service oriented architectures; however, it is still worth pointing out the differences. A service choreography model works without a central orchestrator while a service orchestration model relies on a central controller to couple services. The microservice architectural style has promoted the idea of event-driven architectures to decouple your services. Smart endpoints and dumb pipes are preferred even more now than in previous generations of service-oriented architectures. This has not always been followed in all SOA implementations, resulting in ESB misuse [7].

## Control Flow Modeling

The workflow concept has de-facto become the standard paradigm for process modeling in business process management systems. Workflows are typically looked from three perspectives: 1) the control perspective, describing the logical order of tasks; 2) the data perspective, describing the information exchange between tasks; and 3) the resource perspective, describing the originators of tasks. Industry standards such as UML activity diagrams [8], the Business Process Model and Notation (BPMN) [9] and event-driven process chains [10] offer graphical notations for stepwise processes that include choice, iteration, and concurrent execution. However, data flow and control flow in business workflows are not independent. The routing decisions in a workflow are typically based on data. The emphasis of these workflow modeling approaches is on control flow.

BPMN [9], for instance, does support data objects and data stores, so it is possible to use it to represent data flow, but control-flow modeling dominates BPMN models. Let's take a look at an illustrative example from our Industrial DevOps project [11]. Figure 1 shows an example BPMN workflow for temperature control of engines in a production line.

Timer events in BPMN are events which are triggered by a defined timer, in this example a temperature sensor that periodically measures the engine's temperature. The engine control checks the temperature. If the temperature is too high or too low, the engine receives orders to decrease or increase the temperature, respectively. The measured temperatures are written to the time series database. Corresponding to the temperature check, appropriate messages are written to the log database. In case the temperature is too low or too high, the operator is alerted. The temperature is measured immediately after decelerating / accelerating the engine and periodically triggered by the timer event.

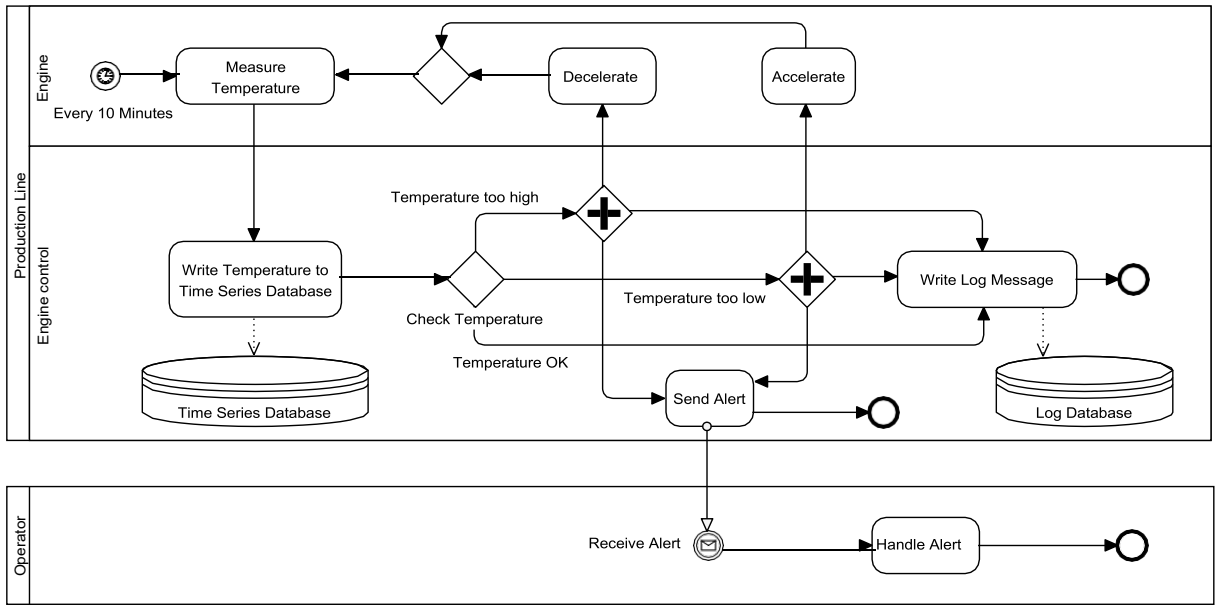

Figure 1: An example BPMN diagram for engine temperature control in a production line.

*Check Temperature* is a data-based exclusive gateway [9]. BPMN offers several other gateways, including inclusive and parallel gateways. If, for instance, the temperature is too high, the production line receives the order to decrease the temperature, the corresponding log message is written, and the operator is alerted, all in parallel.

The graphical, horizontal notation pools in Figure 1 depict the participants within the collaboration. Each pool forms a container for some processes. States in BPMN are linked by sequence, exception or message flows; sequence flows can be either incoming to or outgoing from a state. While *sequence* flows are restricted to an individual pool, *message* flows represent communications between pools. In Figure 1, alerting messages are exchanged between the Production Line and the Operator, depicted as a dashed arrow. In this example, no data objects are modeled, the emphasis is on control flow.

It is not apparent how to map parts of this model to bounded contexts and microservices. According to Evans [6], when a significant process or transformation in the domain is not a natural responsibility of an entity or value object, you should add an operation to the model as interface declared as a domain service. For instance, checking the temperature could be considered such a domain service. However, in Figure 1, this decision is modeled as control flow in the 'central' workflow. As discussed in the previous section, we get a conflict between central control (via orchestration) and independent evolution of microservices when modeling this control flow on the integration level.

Our BPMN engine temperature control example is dominated by modeling the control flow. As an alternative approach, the JOpera Visual Composition Language [12] originates from the workflow area, but the JOpera approach emphasizes data flow. From the JOpera data flow graph, it is possible to derive the process' control flow graph. JOpera includes a separate graph for modeling the control flow to specify control flow dependencies that cannot be automatically derived from the data flow. Before we take a look at such a combination of data and control flow, let's take a look at pure data flow modeling in the following section.

## Data Flow Modeling

A data flow diagram (DFD) is a modeling technique for describing and analyzing information flows. It illustrates the flow of information based on input and output data. DFDs support structured analysis and design. They have the purpose of clarifying system requirements and major data transformations. DFDs illustrate business processes with the help of external data stores, the data flowing from one process to another, and delivering the result data. A DFD is a way to visualize the flow of data of a process or a system that aims to be accessible to both software engineers and domain experts alike. A DFD has no control flow, there are no decision rules and no loops.

Several DFD notations exist. We employ the Gane & Sarson notation [13]. To facilitate the understanding of DFDs, the example DFD for engine temperature control in Figure 2 displays the four basic elements Process, Data Store, Data Flow and External Entity, which are introduced as follows:

- Processes refer to the activities that operate the data of the system. A process receives input data and produces output with a different content or form. Processes can be as simple as collecting input data, or it can be complex as producing a report containing monthly sales. A process is depicted as squares with rounded corners with a unique name in form of verb or verb phrase, for example, Send Temperature in Figure 2. It is optional to indicate the place at which the process is executed, 'Engine Control' in our example.

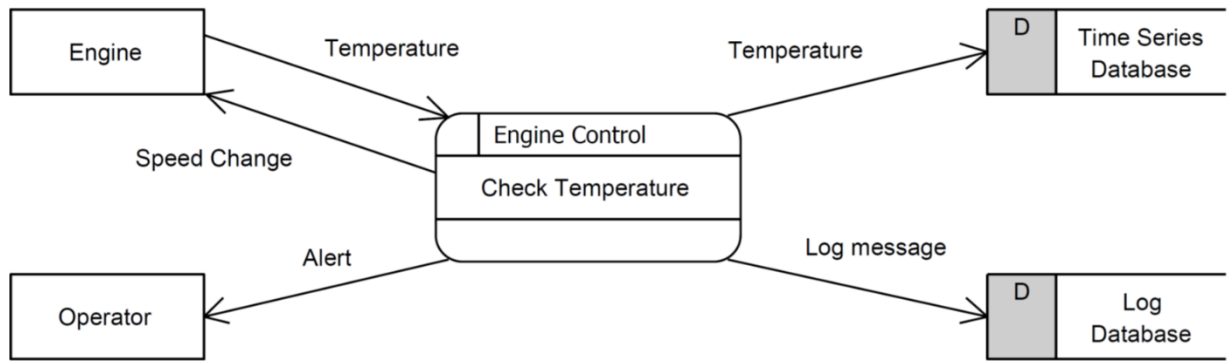

Figure 2: An example DFD for engine temperature control. Processes are depicted as squares with rounded corners, data store as open rectangles, external entities as closed rectangles, and data flows as directed lines.

- Data Stores represent the repository of data manipulated by processes, which can be databases or files (Time Series Database and Log Database in Figure 2). A data store is represented by an open rectangle in a DFD with a name in the form of noun or noun phrase. A data store is used to represent a situation when the system must retain data because one or more processes need to use the stored data in a later time.

  Note that such data stores could also be modeled with the BPMN. Our goal here is to explicitly illustrate the differences between control-flow modeling (with BPMN in Figure 1) and data-flow modeling (with DFD in Figure 2).

- Data Flows are directed lines indicating the data flow from or to a process with the information on the line of a data flow. At least one end of a data flow is linked to a process. Note that data cannot move without a process. In other words, data cannot go to or come from a data store or an external entity without having a process pushing it or pulling it. Data stores are passive while processes and external entities are active. In Figure 2, the Engine sends the measured temperature to the Engine Control to Check the Temperature.

- External entities are components outside of the boundaries of the modeled information system. They represent how the information system interacts with the outside world. Example external entities in Figure 2 are the Engine and the Operator. An external entity is depicted as a closed rectangle in DFDs. An external entity is a person, department, outside organization, or other information system that provides data to the system or receives outputs from the system.

Note that the DFD in Figure 2 does not describe exactly the same information as the control-flow BPMN diagram in Figure 1. Several (control-flow) details are left out, such as the logic to check the temperature. This decision is delegated to the appropriate domain service [6], for which a microservice will be responsible in a microservice architecture (this could be a domain service 'Check Temperature' for our engine control example). Such control-flow concerns are delegated from the integration layer to the individual microservices, to reduce the coupling. Please note that just such a simple rule-based decision as in our illustrative example would be somewhat too fine grained to constitute a microservice.

## Titan Flow-based Programming

Titan is a software platform for integrating and monitoring industrial production environments, following our Industrial DevOps approach [11]. With the Industrial DevOps approach, we intend to introduce methods and culture of DevOps into industrial production environments. The fundamental concept of this approach is a continuous process of development, operation, and observation of the entire production environment.

To achieve this, Titan applies the principles of flow-based programming. Flow-based programming is a programming paradigm, introduced by Morrison in the early 1970s [14]. Flow-based programming defines applications as networks of black-box processes, which communicate via data traveling across predefined connections. As such, flow-based programming emphasize data flows, as the previously introduced DFDs do. Example Titan flows for engine temperature control are shown in Figure 3. In Titan, a graph of connected bricks is called a *flow*. There are several types of bricks [15]:

- Inlets and outlets are to be found at the edges of flows and constitute the start and ending points of data flows (depicted as triangles embedded within squares). An inlet is a producer brick to a data flow. An outlet brick will end a data flow. In Figure 3, Engine is an inlet for temperature measurements. The Time Series Database Writer is an outlet to store the measurement data. Within Titan flows, communication among bricks is standardized, while inlets and outlets connect our flows to possibly heterogeneous external components.

- Filter bricks (depicted as triangles, with the tip to the right) process the incoming data based on filter conditions. The Temperature Filter in Figure 3 converts the raw sensor measurements into corresponding Titan data structures.

- Selector bricks (depicted as triangles, with the tip to the left) forward incoming data to selected out ports of this brick depending on the conditions set in the implementation. In

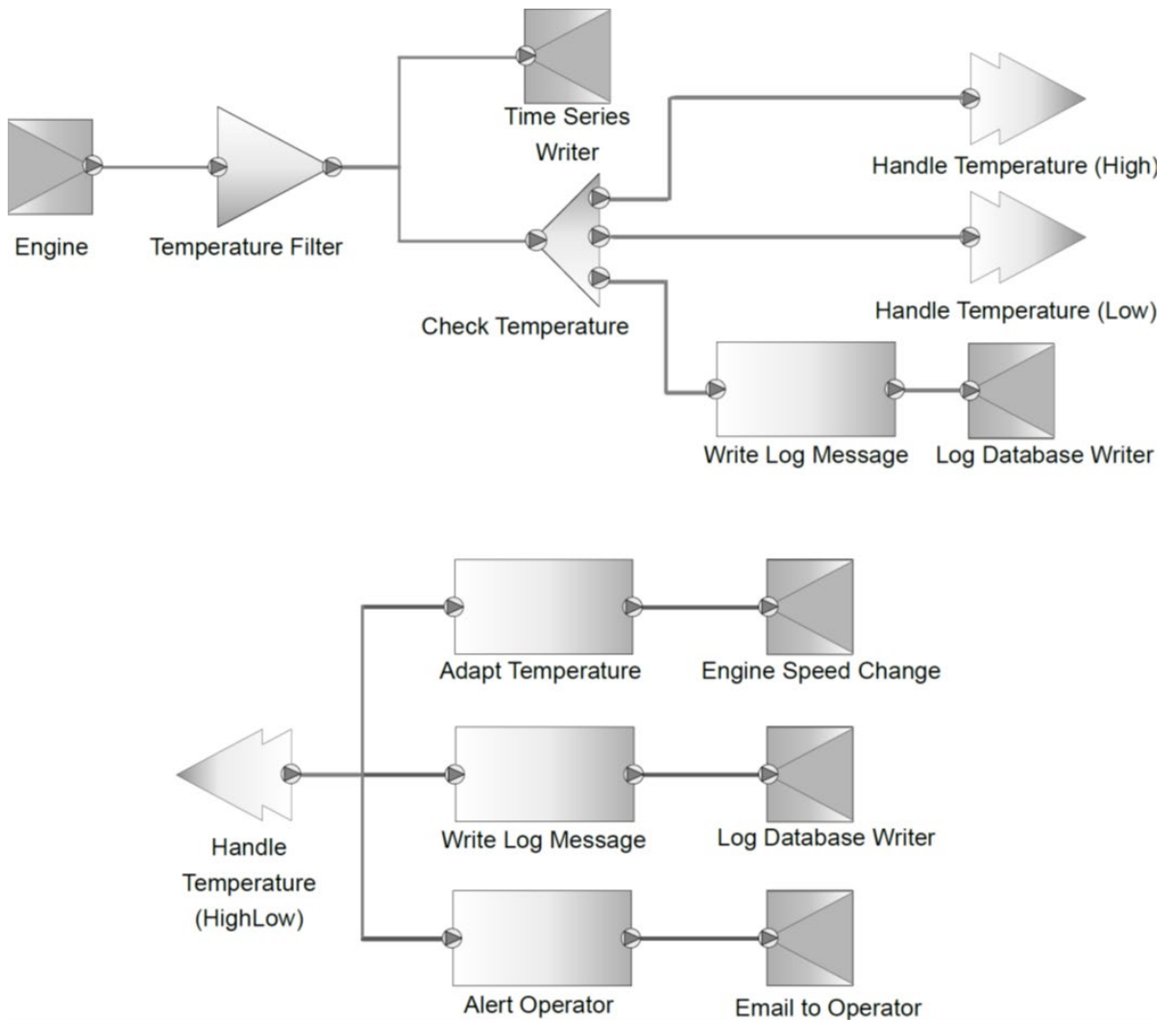

Figure 3: Example Titan flows for engine temperature control.

Figure 3, the Check Temperature selector decides whether to handle the temperature or to write a log message.

- Signals do constitute a special type of flow edge (depicted as double arrows to the left and right for signal producers and consumers, respectively). A signal producer starts a flow by receiving a trigger possibly including a data set delivered via the signal. The signal consumer will end the current flow, optionally triggering some signal producers with the same name. In Figure 3, Handle Temperature is such a signal. With signals, we may build event-driven architectures.

- General bricks (depicted as rectangles) contain logic that cannot be classified into the more specialized bricks listed above. In Figure 3, Adapt Temperature and Alert Operator are examples for such general bricks.

Note that the control flow decisions are encapsulated within the brick implementations. The visual Titan flows describe only data flow. The essential difference to the previous DFD model is that we model *event-driven architectures* with Titan's signal mechanism. Event-driven architectures fit to microservice architectures [16].

Titan provides a graphical modeling language, which is designed to enable the domain experts to model the integration and, based on this, to configure the integrated system. Hence, there is no software engineer required to perform these configurations or (pre-configured) changes. Instead, domain experts receive training on the modeling language, to allow for end-user programming.

The strong encapsulation of the internal functionality of a brick makes it highly modular. Bricks implemented in different languages can be combined. The brick logic can be described in the form of a script, for example with Python. With Titan, we also intend to combine our flow-based programming approach with block-based programming [17], such that the domain experts do not need to learn textual programming languages to describe the internal brick logic. In block-based programming, the programming constructs like conditionals and loops are represented via graphical blocks. Popular examples include MIT Scratch and Google Blockly. Figure 4 depicts a block-based program snippet and corresponding Python code for the Check Temperature selector brick of Figure 3.

With Titan, graphical flow-based programming for integrating distributed systems is dominated by data flows, while control flow is specified within the bricks. For both, we provide low-code programming to domain experts. In principle, the Titan approach is similar to JOpera [12], where data flow graphs are refined by control flow graphs. However, with Titan, data flow graphs are refined via block-based programming to specify the control flow (Figure 4).

## Conclusion

With the production line example, we intend to illustrate our experience and lessons learned with respect to control and data flow in distributed systems integration:

1. Our BPMN engine temperature control example in Figure 1 is a pure control-flow model. The logic for checking the temperature, for instance, is explicitly modeled in the (global) workflow.

2. Our DFD engine temperature control example in Figure 2 is a pure data-flow model. The logic for checking the temperature is *not* modeled in the (global) data flow, it should be implemented in a domain service.

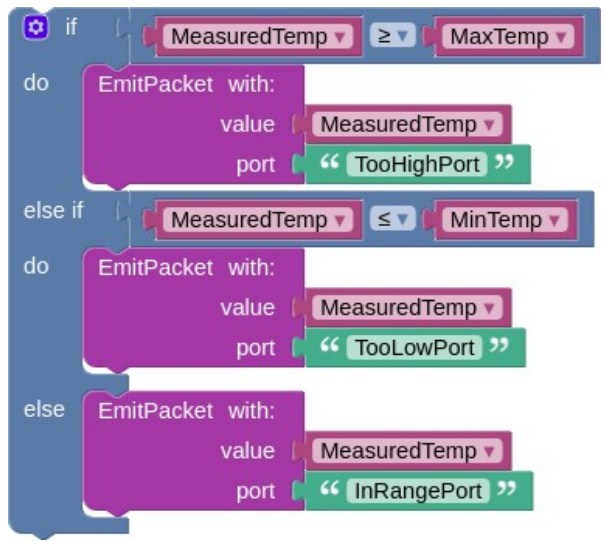

```
if MeasuredTemp >= MaxTemp:
    EmitPacket(MeasuredTemp, 'TooHighPort')
elif MeasuredTemp <= MinTemp:
    EmitPacket(MeasuredTemp, 'TooLowPort')
else:
    EmitPacket(MeasuredTemp, 'InRangePort')
```

Figure 4: Block-based program snippet (left) and corresponding Python code (right) for the Check Temperature selector brick of Figure 3.

3. Our Titan engine temperature control example in Figure 3 is a data-flow model, enriched with events (called signals in Titan). Such event-driven architectures fit well to microservice architectures and domain-driven design [16].

Both, data and control flow are important concerns in distributed systems integration. Based on our experience, we suggest that for loose coupling, explicit control-flow modeling should be avoided on the integration level. Modeling control flow is more coupled because it as- sumes an exact ordering of service invocations, while data flow abstracts from this ordering as long as the service interfaces have matching assumptions regarding the data that needs to be exchanged. Thus, on the level of integrating microservices, data-flow modeling should be dominant. Control-flow should be secondary and preferably delegated to the microservices in some way. However, be aware of the resulting trade-off of loosing an integrated overview on the control flow. Such an integrated overview on the actual system interactions may be reconstructed via runtime monitoring [18, 19], but the system design should focus on data flow. We suggest that researchers investigate more on data-flow oriented modeling methods, and that professionals reconsider and revive data-flow modeling and flow-based programming.

## Acknowledgement

This research is funded by the Federal Ministry of Education and Research (BMBF, Germany) in the Titan project (https://www.industrial-devops.org, grant no. 01IS17084A/B).

**Wilhelm Hasselbring** is a Full Professor of software engineering in the Department of Computer Science at Kiel University, Germany. Contact him at hasselbring@email.uni-kiel.de.

**Maik Wojcieszak** is Co-founder and CTO of wobe-systems GmbH, providing software solutions for industrial automation. Kiel, Germany. Contact him at mw@wobe-systems.com.

**Schahram Dustdar** is a Professor of Computer Science and the Head of the Distributed Systems Group, TU Wien, Vienna, Austria. He became Fellow of the IEEE in 2016 for contributions to elastic computing for cloud applications. He is the corresponding author of this article. Contact him at dustdar@dsg.tuwien.ac.at.